\documentclass[10pt,twocolumn]{article}

\usepackage[letterpaper,margin=0.75in,columnsep=0.28in]{geometry}
\usepackage[T1]{fontenc}
\usepackage{microtype}
\usepackage{times}

\usepackage{amsmath,amssymb}
\usepackage{booktabs}
\usepackage{array}
\usepackage{tikz}
\usetikzlibrary{arrows.meta,positioning,fit,backgrounds,calc,shapes.geometric}
\usepackage{xcolor}
\usepackage{caption}
\usepackage{enumitem}
\setlist{nosep,leftmargin=1.2em}
\usepackage[hidelinks]{hyperref}
\usepackage{url}

\definecolor{hot}{HTML}{C0392B}
\definecolor{live}{HTML}{2E86C1}
\definecolor{snap}{HTML}{8E44AD}
\definecolor{brain}{HTML}{1E8449}
\definecolor{ink}{HTML}{1B2631}
\definecolor{soft}{HTML}{F4F6F7}
\definecolor{edge}{HTML}{5D6D7E}

\usepackage{titlesec}
\titlespacing*{\section}{0pt}{8pt}{3pt}
\titlespacing*{\subsection}{0pt}{6pt}{2pt}
\titleformat{\section}{\normalfont\large\bfseries}{\thesection}{0.6em}{}
\titleformat{\subsection}{\normalfont\normalsize\bfseries}{\thesubsection}{0.6em}{}

\begin{document}

\twocolumn[
\begin{@twocolumnfalse}
\begin{center}
{\LARGE\bfseries OpsCortex: Operational Memory for Self-Diagnosing\\[2pt]
Microservice Systems\par}
\vspace{8pt}
{\large A Topology-Aware, Memory-Centric Architecture that Separates\\
Root-Cause \emph{Derivation} from Root-Cause \emph{Explanation}\par}
\vspace{12pt}
{\normalsize Momil Seedat\par}
\vspace{2pt}
{\small\itshape Independent Research Prototype\par}
\vspace{14pt}
\end{center}
\begin{abstract}
\noindent\small
Modern microservice deployments fail in ways that are easy to \emph{detect} and
hard to \emph{explain}. When a fault propagates along service dependencies, alerts
fire in floods, dashboards multiply, and the scarce resource, an engineer who
understands how the services relate, is consumed reconstructing context that the
monitoring stack discarded. We argue that the missing ingredient in autonomous
operations is not a better anomaly detector or a larger language model, but
\emph{operational memory}: a persistent, structured representation of how a system
normally behaves, how its parts depend on one another, and how it has failed
before. We present \textsc{OpsCortex}, a working multi-agent prototype that
organizes this memory into four tiers and uses it to separate two tasks the field
usually conflates: \emph{deriving} a root cause and \emph{explaining} it. Root
cause is computed deterministically from a learned dependency graph and the
temporal ordering of threshold crossings; a large language model (LLM) is then
asked only to explain, confirm, and recommend, using evidence the system has
already assembled. We motivate the design with two documented production
cascading failures, review representative literature on observability, anomaly
detection, graph-based localization, and LLM-assisted diagnosis, and show how each
architectural choice maps directly to a failure mode those incidents exhibit. The
prototype is validated on an instrumented e-commerce benchmark with eight
injectable failure scenarios.
\vspace{8pt}
\end{abstract}
\end{@twocolumnfalse}
]

\section{Introduction}
A microservice incident is rarely a mystery of \emph{whether} something is wrong.
The error rate crosses a line, latency spikes, a queue backs up: detection is the
easy part, and the industry has largely saturated it. The hard part is everything
after the alert: which of the dozens of services that just turned red is the
\emph{cause} and which are merely \emph{victims} of a cascade; whether a breach is
a genuine incident or the same harmless nightly batch job; and what an on-call
engineer should actually do at 3\,a.m.

Two documented production failures make the problem concrete. In Slack's outage of
4 January 2021, an overloaded AWS Transit Gateway caused packet loss that
cascaded through the serving tier; the company's own report notes that ``a large
part of the challenge was understanding what the core problem was,'' and that the
dashboarding and alerting service \emph{itself} became unavailable during the
incident because it shared a failure dependency with the affected
network~\cite{slack2021}. In Slack's 22 February 2022 incident, a routine
configuration rollout crossed a tipping point at peak traffic and entered what the
authors explicitly call ``a cascading failure scenario'': a textbook example of
complex-systems failure with several interacting contributing
factors~\cite{slack2022}. Both incidents share a signature: detection was prompt,
but \emph{localization and explanation} were slow, manual, and obstructed by the
loss of context.

Three structural problems recur. First, \textbf{alert fatigue}: threshold systems
are memoryless and cannot tell that an alert has fired harmlessly many times, nor
that a value benign on Sunday is a crisis on Wednesday. Second, \textbf{lost
context}: the call graph at the moment of failure, the per-time-of-day notion of
``normal,'' and what changed just beforehand are scattered, transient, or never
recorded, so diagnosis becomes archaeology. Third, the \textbf{cascade problem}:
failure propagates along dependency edges, so the signal an engineer needs is
\emph{temporal and topological} (who broke first, and along which path the damage
spread), yet flat alert streams discard exactly this structure.

We propose a memory-centric response. An operations system should accumulate
\emph{understanding} the way a long-tenured engineer does: not by being cleverer in
the moment, but by remembering. This reframes the goal from \emph{infer the root
cause from scratch on every alert} to \emph{maintain a living model of the system
so that, when an alert arrives, the answer is mostly already known}. Three design
commitments follow, and they organize the rest of this paper:
(i)~memory is the substrate, not an afterthought;
(ii)~\emph{derivation} and \emph{explanation} are different jobs and belong to
different mechanisms; and
(iii)~the system should get cheaper as it learns, reserving the expensive LLM call
for genuinely novel faults.

\noindent\textbf{Contributions.}
\begin{itemize}
\item A four-tier operational-memory architecture (Section~\ref{sec:mem})
separating present state, live topology, change history, and permanent knowledge.
\item A two-stage root-cause method that computes the cause deterministically
(graph BFS + temporal ordering) and uses an LLM strictly as an
\emph{explainer} of pre-computed evidence (Section~\ref{sec:design}).
\item A learned suppress-then-detect lifecycle that silences structural noise yet
overrides suppression on sudden spikes and gradual drift (Section~\ref{sec:design}).
\item An explicit mapping from each design choice to the failure modes of two
documented production incidents (Section~\ref{sec:fit}).
\end{itemize}
The full implementation, benchmark harness, and the eight scenarios are open
source~\cite{repo}.

\section{Literature Review}\label{sec:lit}
Research on diagnosing distributed systems spans observability infrastructure,
anomaly detection, graph-based localization, and, most recently, LLM-assisted
incident management. We organize a compact review around five representative works
that bracket the design space \textsc{OpsCortex} occupies.

\paragraph{Observability foundations.}
The ability to reason about a request as it crosses service boundaries was
established by distributed tracing. Sigelman et al.'s \emph{Dapper}~\cite{dapper}
introduced low-overhead, application-transparent, sampled tracing at Google and
made the point, central to our work, that ``the overall system structure has to
be inferred'' rather than read from a static configuration, because deployments
change continuously. \textsc{OpsCortex} inherits this stance: it hard-codes no
service list and infers topology from telemetry at runtime. Where Dapper provides
the raw causal threads, our system adds a persistent model on top of them.

\paragraph{Anomaly detection.}
Detecting that \emph{something} is unusual is a mature problem. The Isolation
Forest of Liu, Ting, and Zhou~\cite{iforest} isolates anomalies directly via
random partitioning rather than profiling normality, achieving linear time and low
memory, properties that make per-service, multi-metric scoring cheap enough to run
continuously. \textsc{OpsCortex} uses one Isolation Forest per service over the
full metric vector, so that a combination of individually-subthreshold metrics can
still be flagged as collectively unusual. The limitation, shared by all pure
detectors, is that an anomaly score is not an explanation and carries no
remediation; this is precisely the gap our LLM layer is scoped to fill.

\paragraph{Graph-based root-cause localization.}
Because failure propagates along dependencies, a productive line of work localizes
faults on a service graph. MicroRCA~\cite{microrca} builds an \emph{attributed
graph} of anomaly propagation across services and hosts and ranks candidates with a
personalized-PageRank centrality, reporting high precision on injected faults in a
Kubernetes benchmark without application instrumentation. \textsc{OpsCortex} shares
the premise that the dependency graph is the right substrate for localization, but
differs in method: rather than a single centrality score, it uses an explicit,
auditable rule (breadth-first reachability over confirmed call edges, then the
\emph{temporal} order of threshold crossings) so that the cause/victim distinction
is reproducible and inspectable rather than emergent from a ranking.

\paragraph{Surveying the field.}
The breadth of these techniques is captured by Wang and Qi's comprehensive survey
of root-cause analysis in (micro)services~\cite{rcasurvey}, which taxonomizes
metric-, trace-, log-, and multimodal methods and explicitly identifies
graph-theoretic and time-series approaches as dominant families, while flagging
evaluation and the integration of heterogeneous signals as open challenges. Our
design can be read as a deliberate combination of two families the survey treats
separately (time-series baselining for \emph{detection} and graph reasoning for
\emph{localization}), unified by a persistent memory.

\paragraph{LLMs for incident management.}
The newest wave points language models at incident data. Ahmed et al.~\cite{llmrca}
conducted the first large-scale study (40{,}000+ Microsoft incidents) of GPT-class
models for recommending root causes and mitigations, finding genuine promise but
also that models asked to \emph{produce} a root cause from limited context are
uneven and require fine-tuning and human oversight. We take this as evidence for a
division of labour: an LLM that must \emph{derive} causation from raw telemetry is
being asked to do the wrong job. \textsc{OpsCortex} therefore narrows the model's
task to explaining and confirming a cause the deterministic layer has already
identified, with evidence supplied in the prompt, trading open-ended generation
for grounded, lower-variance output.

\smallskip
\noindent In sum, the literature supplies the pieces (tracing for structure,
isolation forests for cheap detection, attributed graphs for localization, LLMs for
natural-language diagnosis) but tends to treat them in isolation. The
contribution of \textsc{OpsCortex} is architectural: a memory that makes detection
time-aware, a deterministic layer that makes localization auditable, and an LLM
confined to the explanatory role it is actually good at.

\section{The Four-Tier Memory}\label{sec:mem}
\begin{figure}[t]
\centering
\begin{tikzpicture}[
  font=\scriptsize,
  tier/.style={rounded corners=2pt, draw=#1, line width=0.9pt, fill=#1!6,
               text width=3.55cm, inner sep=4pt, minimum height=1.05cm},
  lab/.style={font=\scriptsize\bfseries},
  ar/.style={-{Stealth[length=2mm]}, draw=edge, line width=0.7pt},
]
\node[tier=hot] (t1) at (0,0)
  {\textbf{\textcolor{hot}{Tier 1: Redis (hot)}}\\[1pt]
   node:*, timeseries:*, stream:alerts\\
   \emph{lifetime: $\sim$2\,h TTL, the present}};
\node[tier=live, below=0.42cm of t1] (t2)
  {\textbf{\textcolor{live}{Tier 2: NetworkX (live graph)}}\\[1pt]
   3-level topology; \texttt{CALLS} edges; status\\
   \emph{lifetime: process, fast working view}};
\node[tier=snap, below=0.42cm of t2] (t3)
  {\textbf{\textcolor{snap}{Tier 3: Neo4j (snapshots)}}\\[1pt]
   full graph every 60\,s; status progression\\
   \emph{lifetime: permanent, how it changed}};
\node[tier=brain, below=0.42cm of t3] (t4)
  {\textbf{\textcolor{brain}{Tier 4: SQLite (permanent brain)}}\\[1pt]
   baselines, confirmed edges, noise,\\ incident history\\
   \emph{(what is true)}};

\draw[ar] (t1.east) to[bend left=32] node[right,lab,xshift=1pt]{sync 5\,s} (t2.east);
\draw[ar] (t2.west) to[bend left=32] node[left,lab,xshift=-1pt]{snap 60\,s} (t3.west);
\draw[ar, dashed, draw=brain] (t4.east) to[bend left=40]
   node[right,lab,align=left,xshift=1pt]{cold-start\\rebuild} (t2.east);
\draw[ar, dashed, draw=brain] (t3.west) to[bend left=18] (t4.west);
\end{tikzpicture}
\caption{The four memory tiers, ordered by lifetime. Solid arrows are the live
write path; dashed arrows are the permanent store reconstituting the volatile
tiers on restart, so the system is never blind.}
\label{fig:mem}
\end{figure}
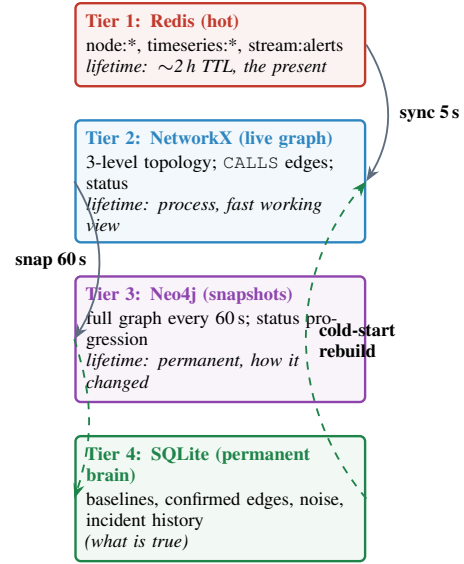

The system's primary artifact is not a model or an agent but its \emph{memory}.
Detection and diagnosis are queries against it. Memory is layered by lifetime and
purpose so that ``what is happening now,'' ``how things connect,'' ``how the system
changed recently,'' and ``what is permanently true'' never compete for the same
store (Figure~\ref{fig:mem}).

\textbf{Tier 1: hot working memory (Redis).} Current metric values, per-metric
sliding windows, and an alert stream, each expiring within hours. This is the
system's present tense.

\textbf{Tier 2: live dependency graph (NetworkX).} A three-level directed graph
(service\,$\rightarrow$\,category\,$\rightarrow$\,metric) plus \texttt{CALLS} edges,
held in process for fast traversal. Status propagates upward: worst metric sets the
category, worst category sets the service.

\textbf{Tier 3: snapshot history (Neo4j).} The full graph is serialized every
60\,s. This lets the system ask not only \emph{what is the state} but \emph{how did
it change} in the minutes before an incident, the signal that distinguishes a slow
leak from a sudden crash.

\textbf{Tier 4: permanent brain (SQLite).} Everything learned: per-(service,
metric, hour, weekday) baselines; confirmed call edges with a confidence count that
grows through repeated observation; a registry of structural noise; and the full
history of past diagnoses. On restart this tier reconstructs the graph and
baselines \emph{before} any new telemetry arrives.

\section{Design: Derive, Then Explain}\label{sec:design}
\begin{figure}[t]
\centering
\begin{tikzpicture}[
  font=\scriptsize,
  node distance=0.5cm,
  box/.style={rounded corners=2pt, draw=edge, line width=0.8pt, fill=soft,
              text width=4.6cm, align=center, inner sep=4pt, minimum height=0.72cm},
  acc/.style={rounded corners=2pt, draw=brain, line width=1pt, fill=brain!7,
              text width=4.6cm, align=center, inner sep=4pt, minimum height=0.72cm},
  llmb/.style={rounded corners=2pt, draw=snap, line width=1pt, fill=snap!7,
              text width=4.6cm, align=center, inner sep=4pt, minimum height=0.72cm},
  ar/.style={-{Stealth[length=2mm]}, draw=edge, line width=0.8pt},
]
\node[box] (svc) {Services \emph{(OTel traces + /metrics)}};
\node[box, below=of svc] (mon)
   {Monitoring agent \;$\rightarrow$\; Redis alert stream};
\node[acc, below=of mon] (topo)
   {Topology agent: live graph + ML brain\\
    \emph{baselines, isolation forest, noise, drift}};
\node[acc, below=of topo] (rc)
   {\textbf{Derive (deterministic)}\\
    BFS over confirmed edges $+$ temporal order};
\node[llmb, below=of rc] (llm)
   {\textbf{Explain (generative)}\\
    LLM confirms cause, recommends fix};
\node[box, below=of llm] (inc)
   {Incident store (SQLite) $+$ dashboard};

\draw[ar] (svc) -- (mon);
\draw[ar] (mon) -- (topo);
\draw[ar] (topo) -- node[right=1pt,font=\scriptsize]{graph $+$ scores} (rc);
\draw[ar] (rc) -- node[right=1pt,font=\scriptsize]{cause $+$ evidence} (llm);
\draw[ar] (llm) -- (inc);
\draw[ar, dashed, draw=snap, line width=0.9pt]
   (rc.west) -- ++(-0.55,0)
   node[midway,left=1pt,font=\scriptsize,align=center,text width=1.6cm]
   {pattern hit:\\ skip LLM}
   |- (inc.west);
\end{tikzpicture}
\caption{The diagnosis pipeline. The deterministic stage (green) derives the root
cause and assembles evidence; the generative stage (purple) only explains and
recommends. On a strong match to a past incident, the pattern matcher writes the
diagnosis directly and bypasses the LLM (dashed).}
\label{fig:flow}
\end{figure}
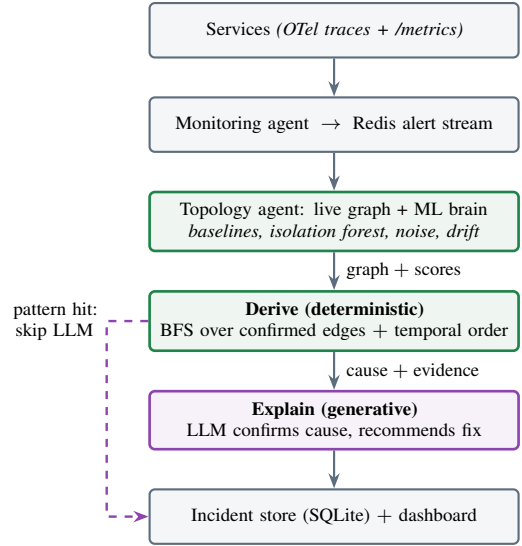

Figure~\ref{fig:flow} shows the pipeline. Three always-on agents observe the
system; the diagnosis path is deliberately split into a deterministic derivation
stage and a generative explanation stage.

\subsection{Agents and cold start}
A \emph{monitoring} agent polls metrics and writes hot state and an alert stream to
Tier~1, accepting any service label it sees so that no allow-list is required. A
\emph{topology} agent rebuilds the live graph from hot state every few seconds,
maintains the ML brain, reconciles a service lifecycle (active, offline,
deprecated), and snapshots the graph to Tier~3. A \emph{diagnosis} agent consumes
alerts, applies noise suppression, and runs the derive-then-explain path. On a cold
start the order matters: the permanent brain (Tier~4) is read \emph{first}, so the
graph topology and per-bucket baselines exist before any live telemetry arrives;
only then do hot-state keys repopulate the graph and the alert stream begin to
flow. The system therefore never starts blind, and a restart mid-incident does not
erase what it had learned, directly addressing the self-dependency that disabled
Slack's dashboards in 2021~\cite{slack2021}.

\subsection{Deterministic root cause}
Root cause is computed, not guessed. A breadth-first walk over \emph{confirmed}
call edges identifies every service connected to the fault. Temporal ordering,
which service crossed its alert threshold first, separates the cause
from its downstream victims:
\[
\text{root} \;=\; \arg\min_{s \in \mathcal{R}(f)} \; t_{\text{amber}}(s),
\]
where $\mathcal{R}(f)$ is the set reachable from faulted service $f$ over
\texttt{CALLS} edges and $t_{\text{amber}}(s)$ is the first threshold-crossing time
of $s$. The output, an inferred root cause and an ordered fault path, is
reproducible and inspectable before any model is invoked.

\subsection{Time-aware baselines}
Detection is made context-sensitive by bucketing baselines per
$(\text{service}, \text{metric}, \text{hour}, \text{weekday})$. A self
$z$-score flags sudden deviation from the norm \emph{for this hour}; because a
rolling mean chases a slow trend and hides it, a second \emph{trend} $z$-score fits
a short linear regression and normalizes the slope by residual spread, catching
gradual degradation. Both are passed forward as override signals.

\subsection{Suppress-then-detect lifecycle}
The most illustrative behaviour is noise handling. A nightly batch job that runs
harmlessly will, on its first few appearances, generate incidents, because the
system has not yet learned it is normal. After enough observations in the same time
bucket it is recognized as structural noise and suppressed. Crucially, suppression
is not absolute: a sudden spike (self $z$-score) or a consistent upward drift (trend
$z$-score), or a correlated anomaly in a peer service, each \emph{overrides}
suppression. The same job is silenced when it behaves and immediately un-silenced
the moment it genuinely degrades.

\subsection{LLM as explainer}
Only after derivation is an LLM invoked, and its prompt \emph{leads} with the
assembled evidence: the inferred root cause, the fault path, the strongest metric
$z$-scores, the Isolation-Forest score, and the per-snapshot status progression.
Its task is narrowed from \emph{find the cause} to \emph{explain and confirm this
cause, challenge it if signals contradict, and recommend a fix}, with confidence
constrained by explicit rules tied to signal agreement. A pattern matcher first
scans recent incidents for a matching (service, metric, value) signature; on a
strong match it reconstructs the prior diagnosis and skips the model entirely, so
cost falls as the system learns.

\section{Does the Design Match the Problem?}\label{sec:fit}
We close the loop by mapping each component back to the failure modes the two
documented incidents exhibit (Table~\ref{tab:fit}).

\begin{table}[t]
\centering\small
\caption{Mapping documented failure modes to design responses.}
\label{tab:fit}
\begin{tabular}{@{}p{2.65cm} p{4.65cm}@{}}
\toprule
\textbf{Failure mode (observed)} & \textbf{OpsCortex response} \\
\midrule
Cause hard to find in a cascade~\cite{slack2021} &
Deterministic BFS + temporal ordering yields an inspectable cause/victim path. \\[2pt]
Monitoring/dashboards share a failure dependency~\cite{slack2021} &
Permanent brain (Tier 4) reconstructs state on cold start; diagnosis does not
depend on live dashboards. \\[2pt]
Tipping point at peak traffic~\cite{slack2022} &
Hour/weekday baselines + trend $z$-score detect drift toward a threshold rather
than only the breach. \\[2pt]
Several interacting factors~\cite{slack2022} &
Per-service Isolation Forest over the full metric vector flags collectively
unusual combinations; call-pattern $z$-scores separate a broken service from an
overloaded caller. \\[2pt]
Alert noise obscures signal &
Suppress-then-detect lifecycle silences structural noise but overrides on spike,
drift, or peer correlation. \\
\bottomrule
\end{tabular}
\end{table}

The correspondence is intentional: the deterministic layer addresses the
\emph{localization} difficulty both reports emphasize; the permanent brain
addresses the \emph{lost-context} and self-dependency problem that made Slack's
2021 dashboards fail when most needed; and the time-aware baselines address the
\emph{tipping-point} dynamic of the 2022 incident, where individually unremarkable
steps crossed a threshold at peak load.

\subsection{Validation status and honest limits}
\textsc{OpsCortex} is validated on an instrumented four-service e-commerce
benchmark with eight injectable scenarios (Table~\ref{tab:scen}) that exercise
overload, cascade, queueing, resource exhaustion, and, most importantly, the
learning behaviours: a known pattern that should bypass the LLM, a batch job whose
benign runs should become suppressed, and a degradation that must defeat that
suppression.\footnote{The implementation and benchmark are available at
\url{https://github.com/momil-seedat/ops-cortex}.} This demonstrates the mechanism
end-to-end but is \emph{not} an evaluation on production data, and we make no
comparative accuracy claims against MicroRCA~\cite{microrca} or LLM-only
baselines~\cite{llmrca}.

\begin{table}[t]
\centering\small
\caption{Injectable scenarios in the benchmark and the behaviour each exercises.}
\label{tab:scen}
\begin{tabular}{@{}p{0.35cm} p{3.1cm} p{3.55cm}@{}}
\toprule
\textbf{\#} & \textbf{Scenario} & \textbf{Behaviour under test} \\
\midrule
1 & Request surge & Threshold breach, single-service overload \\
2 & Queue / consumer lag & Backpressure detection \\
3 & Cascade failure & Cause vs.\ victim via temporal order \\
4 & Known-pattern replay & Pattern matcher bypasses the LLM \\
5 & Internal bottleneck & Resource-exhaustion localization \\
6 & New-service registration & Runtime discovery, lifecycle \\
7 & Batch suppression learning & Benign repeats become suppressed \\
8 & Batch degradation & Drift override defeats suppression \\
\bottomrule
\end{tabular}
\end{table}

Three limits are explicit. (1)~``Who crossed threshold first'' is a proxy for
causation that near-simultaneous or feedback-coupled failures can defeat.
(2)~The absence of labelled, real incident data, a difficulty the
survey~\cite{rcasurvey} underscores, means external validity is unproven.
(3)~A grounded prompt reduces but does not eliminate the risk of a confidently
wrong LLM explanation; surfacing disagreement between the deterministic and
generative stages, rather than smoothing it, is future work.

\section{Future Directions}\label{sec:future}
Four directions would turn the prototype into a defensible scientific result.
\textbf{From temporal heuristics to causal inference.} The first-crossing rule is a
useful but coarse proxy; a principled treatment would adopt causal-discovery or
intervention models to distinguish genuine precedence from coincidence and to
attach a confidence to the causal claim, complementing the centrality-based
ranking of graph methods such as MicroRCA~\cite{microrca}.
\textbf{Evaluation on real corpora.} Credible results need either a benchmark with
ground-truth causal structure or a labelled postmortem corpus, scored not only on
naming the right service but on whether the \emph{explanation} and
\emph{recommendation} were correct, the harder bar identified for LLM-based
diagnosis~\cite{llmrca}.
\textbf{Cross-system transfer.} The memory here learns one deployment deeply;
consolidating experience into knowledge transferable across topologies is the
lifelong-learning question applied to operations.
\textbf{Runtime dependency and change-impact analysis.} The fault path is, in
effect, dependency analysis performed on \emph{runtime} artifacts (services,
metrics, and the call edges between them) rather than on source code. A deployment
or configuration change is a perturbation whose blast radius propagates along the
same edges a cascade does; treating change-impact and incident-cascade as two faces
of one dependency structure, atop the trace-derived topology that
Dapper~\cite{dapper} first made observable, is a promising unification.

\section{Conclusion}
The bottleneck in autonomous operations is not perception but memory. Systems that
detect failure are abundant; systems that \emph{understand} it, that know what is
normal for the moment, how the parts depend on one another, and what has gone wrong
before, are not. \textsc{OpsCortex} is a concrete argument that the way forward is
to make that understanding the central artifact: compute what can be computed
deterministically, reserve generative reasoning for what genuinely needs it, and
let a persistent operational memory make the whole system cheaper, calmer, and more
confident the longer it runs. The prototype shows the shape of the idea; evaluation
on real incident corpora and a causal treatment of precedence are the work of
making it true.


\end{document}